\documentstyle[pasjskel,graphicx,natbib,twocolumn]{article}

\begin{document}

\title{Discovery of Extremely Large-Amplitude Quasi-Periodic Photometric
Variability in WC9-Type Wolf-Rayet Binary, WR 104}

\author{Taichi \textsc{Kato}}
\affil{Department of Astronomy, Kyoto University,
       Sakyou-ku, Kyoto 606-8502}
\email{tkato@kusastro.kyoto-u.ac.jp}

\author{Katsumi \textsc{Haseda}}
\affil{Variable Star Observers League in Japan (VSOLJ), 2-7-10 Fujimidai,
       Toyohashi City, Aichi 441-8135}
\email{khaseda@mx1.tees.ne.jp}

\author{Hitoshi \textsc{Yamaoka}}
\affil{Faculty of Science, Kyushu University, Fukuoka 810-8560}
\email{yamaoka@rc.kyushu-u.ac.jp}

\email{\rm{and}}

\author{Kesao \textsc{Takamizawa}}
\affil{Variable Star Observers League in Japan (VSOLJ), 65-1 Oohinata,
       Saku-machi, Nagano 384-0502}
\email{k-takamizawa@nifty.ne.jp}

\begin{abstract}
  We discovered that the Wolf-Rayet (WR)+OB star binary, WR 104,
renowned for its associated ``dusty pinwheel nebula" recently spatially
resolved with infrared interferometry, exhibits strong quasi-periodic
optical variations with a full amplitude of 2.7 mag.
Such a large-amplitude, continuous variation has been unprecedented
in a WR star.  The optical quasi-period ($\sim$241 d) is
in almost perfect agreement with the interferometric period (243.5$\pm$3 d).
The remarkable agreement of the dominant period in optical variability
with the orbital period supports that the strongly varying dust obscuration
is physically related to the binary motion, rather than sporadic
dust-forming episodes.  Considering the low orbital inclination
(11$\pm$7$^{\circ}$) and the nearly circular orbit inferred from the
interferometric observations, the strongly variable line-of-sight extinction
suggests that the highly structured extinction can be being formed via an
ejection of dust in the direction of the binary rotation axis.
Another viable explanation is that the three-dimensional structure of
the shock front, itself is the obscuring body.
Depending on the geometry, the dusty shock front near the conjunction
phase of the binary can completely obscure the inner WR-star wind and
the OB star, which can explain the amplitude of optical fading
and the past observation of remarkable spectral variation.
\end{abstract}


\begin{figure*}
  \begin{center}
    \FigureFile(140mm,70mm){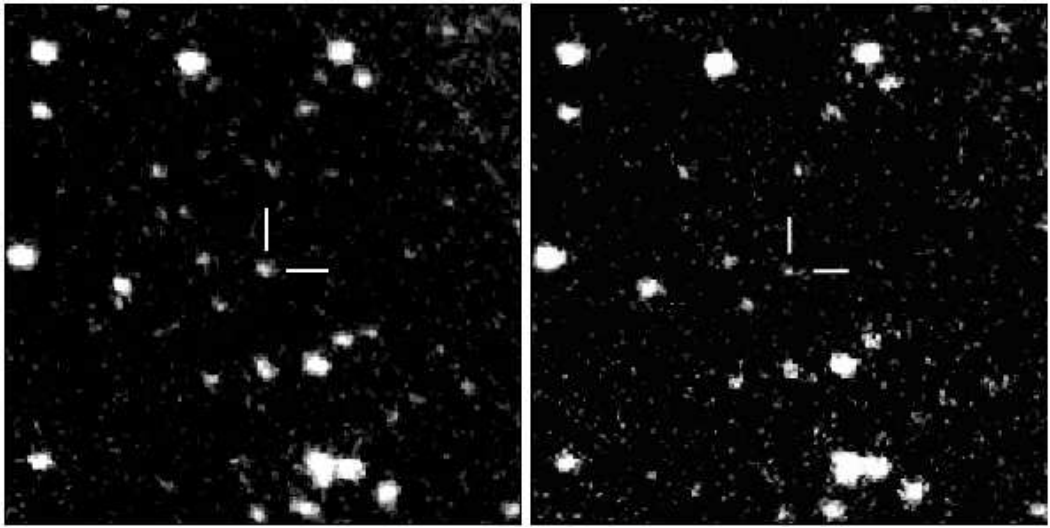}
  \end{center}
  \caption{The variation of WR 104 = HadV82, recorded with photographs
  taken by one of the authors (KH).  The left panel was taken on
  1998 March 2, when the object was at 12.0 magnitude.  The right panel
  was taken on 1998 April 25, when the object was at 13.8 magnitude.
  Such dramatic variability of a Wolf-Rayet star is quite exceptional.
  }
  \label{fig:image}
\end{figure*}

\section{Introduction}
Wolf-Rayet (WR) stars are massive, luminous, hot stars which have blown
away the hydrogen envelope, and are considered to be immediate precursors
of some kinds of supernovae.  Carbon-rich, late-type subclass of WR stars
(WCL) are known as one of the most efficiently dust-productive stellar
environments (for recent reviews,
see \authorcite{wil95} 1995,1997ab).
The dust production in these stars not only highlights the problems of
the dust formation mechanism in a very hot environment \citep{wil97}, but
also plays an important role in the chemical enrichment in galaxies
(\cite{est95}; \cite{sch98}).
In addition to the strong infrared emission in WCL stars, which is
considered to arise from persistent dust shells, there has been emerging
evidence of episodic dust formation in these stars.  \citet{veen97}
reported transient optical fadings in several WCL stars, attributable
to temporary condensations of dust clouds.

\citet{cro97} reported remarkable weakening of the WR-type spectral feature,
together with a possible 1.1 magnitude photometric fading, in a WR+OB
binary WR 104, which was interpreted as a temporary obscuration of the
inner Wolf-Rayet wind by a dust cloud.  On the other hand, recent
interferometric imaging of WR 104 by \citet{tut99} revealed the amazing
``dusty pinwheel nebula".  This extended structure which rotates
with a period of 220$\pm$30 d [\citet{tut02} further reported a refined
the period to be 243.5$\pm$3 d] is believed to be formed by the colliding
winds of the WR star and companion OB-star.  The relation between
occasionally reported obscuration episodes and the persistent dusty
pinwheel structure, however, remained a mystery \authorcite{tut99}
(1999,2002)\footnote{
  More recently, WR 98a and perhaps WR 112 are known to have a
  ``dusty pinwheel nebula" (\cite{mon99}; \cite{mar02}), though neither
  of them are yet known to be as highly variable as WR 104.
}.

In 2001 April, one of the authors (KH) serendipitously discovered a new
variable star named HadV82 \citep{has01}, which was subsequently
identified with WR 104.  This led to a discovery of dramatic photometric
variation in WR 104, with an unprecedented amplitude and frequency of
fadings among all WR stars.
The variation has a quasi-period close to the reported binary period,
which provides an essential clue to understanding the relation between
the formation of dust and the role of binarity.

\section{Observation and results}
A total of 176 observations were done between 1994 May 12 and 2001 April 26,
with twin patrol cameras equipped with D=10cm f/4.0 telephoto lens and
unfiltered T-Max 400 emulsions, located at two sites in Toyohashi, Aichi
(KH) and Saku, Nagano (KT).  The passband of observations covers the range
of 400--650 nm.  Photographic photometry was done using neighboring
comparison stars whose $V$-magnitudes were calibrated by T. Watanabe.
The overall uncertainty of the calibration and individual photometric
estimates is 0.2--0.3 mag, which will not affect the following analysis.

\begin{figure*}
  \begin{center}
    \FigureFile(140mm,70mm){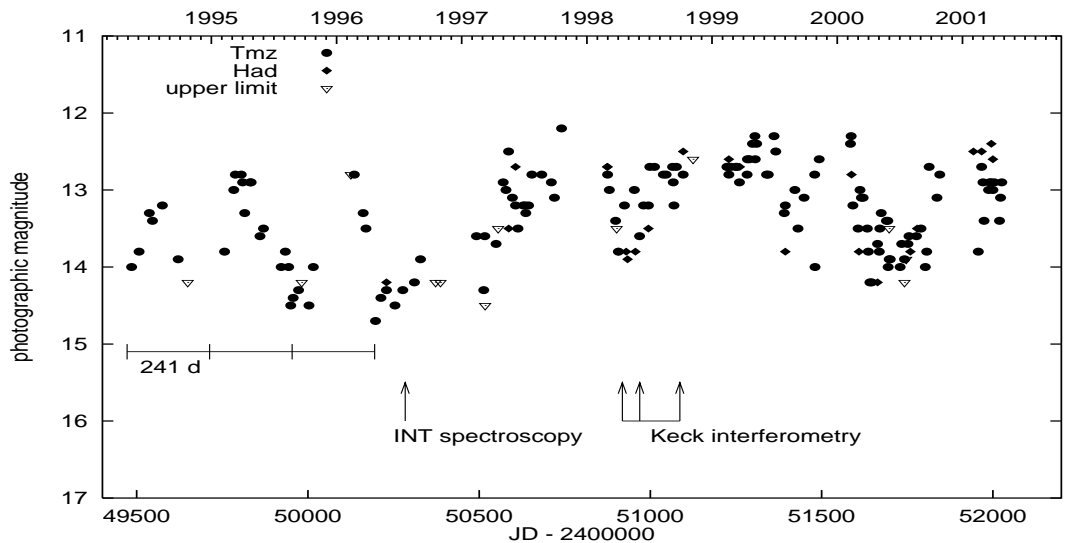}
  \end{center}
  \caption{Light curve of WR 104.  Filled circles and squares represent
  observations by Takamizawa (Tmz) and Haseda (Had) , respectively.
  Open triangles represent upper limits.
  The epochs of the INT spectroscopy \citep{cro97},
  when the weakening of the WR-type spectral feature was observed, and
  Keck interferometry \citep{tut99}, which led to the discovery of the
  ``dusty pinwheel nebula", are marked with arrows.  The coincidence of
  the the remarkable weakening of the WR-type spectral feature and the
  optical fading supports that the WR-type star was almost entirely
  obscured at the time of the observation, and that the WR-type star
  emits most of the visual light.  The 241-d cycle was prominently seen
  for the period of 1994--1996, as shown with a tick-marked line.
  (The tick marks correspond to the phase zero in \citet{tut02}).
  The object slightly brightened in
  1997--1999, when the rigid periodicity became less marked.  Recurring
  fading episodes with time scales of 200--400 d, however, persisted
  up to 2001.  The light curve is completely unique among all previously
  known WR-type variable stars \citep{veen97}.}
  \label{fig:lc}
\end{figure*}

The resultant light curve is presented in figure \ref{fig:lc}.  The star
showed an overall range of variability between 11.8 and fainter than 14.5
magnitudes, making the full amplitude greater than 2.7 magnitudes.
The raw data are available at
$\langle$ ftp://vsnet.kusastro.kyoto-u.ac.jp/pub/vsnet/WR/WR104/wr104obs.jd
$\rangle$.

\begin{figure}
  \begin{center}
    \FigureFile(70mm,60mm){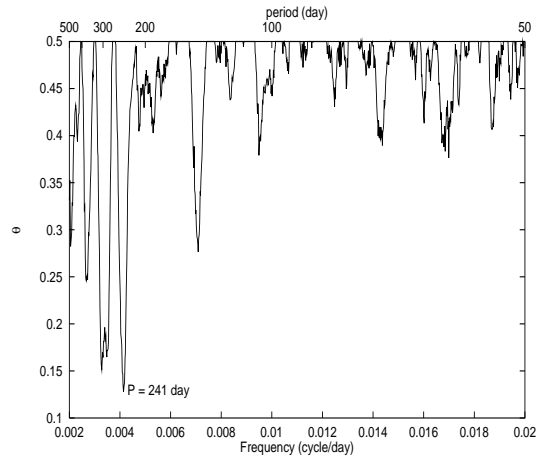}
  \end{center}
  \caption{Period analysis.  Lower theta values represent stronger
  periodicities.  The period of 241 d is the strongest period between
  50 and 500 d (please note that some of the weaker signals are sidelobes
  due to sampling).  The period almost perfectly agrees with the period
  determined with interferometry \citep{tut02}.  This agreement suggests
  that the semi-periodic obscuration of the WR-type component is
  strongly associated with the binary motion.}
  \label{fig:pdm}
\end{figure}

\begin{figure}
  \begin{center}
    \FigureFile(70mm,60mm){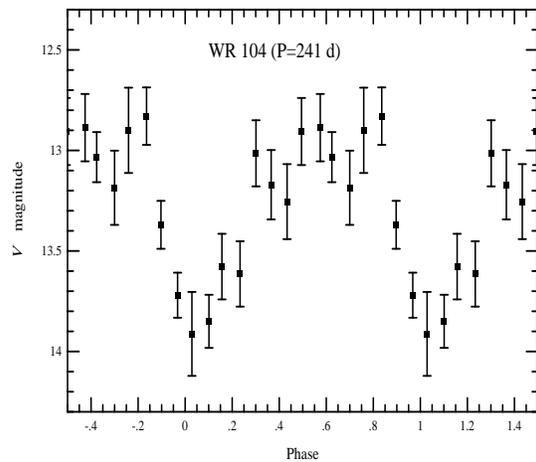}
  \end{center}
  \caption{Phase-averaged light curve of WR 104.  The phase zero corresponds
  to the phase zero in \citet{tut02} (1998 April 14).  The mean orbital
  light curve is characterized by a rather sharp minimum and a broader
  maximum.}
  \label{fig:phase}
\end{figure}

Such a large variation in visible light far exceeds previously known
variations of WR-type stars \citep{veen97}, which only occasionally show
temporary fadings having depths less than 1 magnitude.  Faint phases of
WR 104 typically last an order of months, which is strikingly longer
than any known transient episodes of WR-type variable stars
\citep{veen97}, which usually last days to weeks.
The most remarkable feature is the existence of a quasi-period of
200--400 d, superimposed on a long-term trend, which may be attributed
to slow changes of optical depth of the line-of-sight extinction.
The result of period analysis using the Phase Dispersion Minimization
technique \citep{ste78} is presented in figure \ref{fig:pdm}.
The strongest period in the range of 50--500 d is 241 d, which is in
almost perfect agreement with the suggested orbital period of 243.5$\pm$3 d
\citep{tut02}.  Figure \ref{fig:phase} shows a phase-averaged light curve
with this period.  The mean orbital light curve is characterized by
a rather sharp minimum and a broader maximum.

\section{Discussion}
The previously known most striking evidence of variability in WR 104
is the temporary remarkable weakening of the spectroscopic feature
of the WR-type star \citep{cro97}.  \citet{cro97} suggested that occultation
by a dust cloud condensation, analogous to R CrB stars, is responsible
for the phenomenon.  The coincidence of the the remarkable weakening of
the WR-type spectral feature and the optical fading supports the idea
that the WR-type star was almost entirely obscured at the time of the
observation by \citet{cro97}.

The remarkable agreement of the dominant period in optical variability
with the orbital period supports that at least a considerable fraction
of varying dust obscuration is physically related to the binary motion.
We mainly discuss on implications of the remarkable coincidence between
the period of large-amplitude optical variation and the binary period.

First, the reported low binary inclination of 11$\pm$7$^{\circ}$
\citep{tut02} makes unlikely that the tail of the dust spiral is directly
responsible for the variation.
A possible explanation is the periodic enhancement of dust production in
the WR-star wind at the passage of the companion star though an elliptical
orbit.  This effect is most pronounced in an episodic dust producer, WR 140
\citep{wil90} and presumably WR 137 \citep{wil01}.
The evidence for this scenario in WR 104 is less convincing,
in its little infrared variability \citep{vdh96}, and in its continuous
appearance of the dust spiral \authorcite{tut99} (1999,2002), and the
lack of evidence of a large orbital eccentricity \citep{tut02}, all of
which were considered against episodic dust production around periastron.
Further full-orbit interferometric observation, together with
contemporaneous visual and infrared photometry will be key information in
testing the hypothesis.

The other possibility is that the varying obscuration is of geometrical
origin, e.g. varying extinction as a consequence of rotating binary seen
through a gradient of absorption near the line of sight.  This explanation
is similar to an idea to explain the unique variability of the binary
central star of the planetary nebula, NGC 2346 (\cite{roth84};
\cite{cos86}).  This explanation would require an extremely strong
gradient or a sharply defined dense obscuring body in the line of sight.
The observed amplitude (more than 2.7 magnitudes) could only be explained
if the WR-type component emits about 90\% of the visual light,
and the star is almost completely eclipsed by this obscuring structure.
However, this assumption may not be consistent with its weaker WR-type
spectral feature even in the high state than in other
WC9 stars \citep{coh75}.  Such apparent discrepancies
were also observed in WR 137 \citep{wil01}, which may be a rather common
phenomenon seen in WCL+OB wind collision binaries.

Considering the total line-of-sight absorption of $A_{\rm V}$=6.5 magnitude
\citep{pen94}, about 4.5 mag of which (after subtracting the interstellar
absorption) is likely to be attributed to past mass-loss events of the
progenitor, rather than the present dust formation (\cite{coh75},
\cite{tut99}), such a large fraction (2.7/4.5) of
variable line-of-sight extinction seems to be difficult to reconcile
with the past, presumably more spherical mass-loss episodes.
The present discovery of strongly
variable extinction may alternately suggest that the highly structured
extinction can be being formed via an ejection of dust in the direction
of the binary rotation axis.  Such a possibility may be tested by a
future development of full three-dimensional treatment of dust formation
and ejection in a colliding-wind binary.

Another viable explanation is that the three-dimensional structure of
the shock front, where the compressed gas is considered to effectively
form dust grains \citep{uso91}, itself is the obscuring body.
Depending on the geometry, the dusty shock front near the conjunction
phase of the binary can completely obscure the inner WR-star wind,
as well as the OB star, both are required to explain the spectral
variation \citep{cro97} and the observed amplitude of optical fading.
This explanation would require a higher inclination angle than was
reported, or a sufficient vertical structure to obscure the stellar
component viewed even nearly pole-on.
Detailed high-resolution, full-orbit observation is again indispensable
to discriminate the possibilities.

The present discovery of quasi-periodic, highly variable visual obscuration
in WR 104 proposes a new class and mechanism of light variability
in WR-type stars.  With the advent of sub-milliarcsecond optical and infrared
interferometry, and the coming era of Atacama Large Millimeter Array (ALMA),
WR 104 would provide a powerful tool in geometrically resolving the heart
of dust production in the most efficiently dust-productive stellar
environments.

\vskip 3mm

This work is partly supported by a grant-in-aid (13640239) from the
Japanese Ministry of Education, Culture, Sports, Science and Technology.

\end{document}